# Click A, Buy B: Rethinking Conversion Attribution in E-Commerce Recommendations

Xiangyu Zeng
Meta Platforms, Inc.
Menlo Park, CA, USA
xiangyuzeng@meta.com

Amit Jaspal
Meta Platforms, Inc.
Menlo Park, CA, USA
ajaspal@meta.com

Bin Liu
Meta Platforms, Inc.
Menlo Park, CA, USA
binliubinliu@meta.com

Goutham Panneeru
Meta Platforms, Inc.
Menlo Park, CA, USA
gpanneeru @meta.com

Kevin Huang
Meta Platforms, Inc.
Menlo Park, CA, USA
kevh@meta.com

Nicolas Bievre
Meta Platforms, Inc.
Menlo Park, CA, USA
nbievre@meta.com

Mohit Jaggi
Meta Platforms, Inc.
Menlo Park, CA, USA
mohitjaggi@meta.com

Prathap Maniraju
Meta Platforms, Inc.
Bellevue, WA, USA
prathampm@meta.com

Ankur Jain
Meta Platforms, Inc.
Menlo Park, CA, USA
ankurjain@meta.com

## ABSTRACT

User journeys in e-commerce routinely violate the one-to-one assumption that a clicked item on an advertising platform is the same item later purchased on merchant's website/app. For significant number of converting sessions on our platform, users click product A but buy product B—the Click A, Buy B (CABB) phenomenon. Training recommendation models on raw click-conversion pairs therefore rewards items that merely correlate with purchases, leading to biased learning and sub-optimal conversion rates. We reframe conversion prediction as a multi-task problem with separate heads for Click A → Buy A (CABA) and Click A → Buy B (CABB). To isolate informative CABB conversions from unrelated CABB conversions, we introduce a taxonomy-aware collaborative filtering weighting scheme where each product is first mapped to a leaf node in a product taxonomy, and a category-to-category similarity matrix is learned from large-scale co-engagement logs. This weighting amplifies pairs that reflect genuine substitutable or complementary relations while down-weighting coincidental cross-category purchases. Offline evaluation on e-commerce sessions reduces normalized entropy by 13.9 % versus a last click attribution baseline. An online A/B test on live traffic shows +0.25% gains in the primary business metric.

## 1 Introduction

In e-commerce recommendation systems, a common assumption is that if a user clicks on a product on an advertising platform and later converts i.e. makes a purchase on the merchant's website/app, then that conversion is attributed to that same clicked product. However, user behavior often violates this assumption. A user may click Product A but eventually purchase a different Product B – a scenario we refer to as the Click A, Buy B (CABB) problem (Figure 1). These CABB scenarios are not rare in sessions where users have multiple shopping intent and advertisers' catalogs have very large volumes of items. In fact, on our platform we find that a significant number of conversions exhibit the CABB phenomenon. Naively combining CABA and CABB conversions while building conversion models can lead to suboptimal predictions. If a model is trained assuming all conversions happen on the last clicked item, it will learn to overvalue a subset of highly popular and click-inducing products. As noted in prior work on attribution [1], giving all credit to the last touched item or event can introduce a bias towards rewarding clicks that are simply correlated with conversions, but that might not be causing them. Further naively modeling CABB conversions impacts user experience as well. A recommender that doesn't account for CABB may stop showing certain helpful reference items, reducing the diversity of recommendations. Users might then miss out on discovering products that could be useful. In summary, the CABB problem leads to biased learning and suboptimal recommendations.

To address these issues, we propose a novel approach that reframes recommendation conversion prediction as a multi-task learning problem encompassing both direct and cross-item conversions. By explicitly modeling CABB events, we aim to de-bias the learning process and capture true user intent. Our method leverages similarity between products to discern when a click on A that led to purchase of B was likely a purposeful substitution (e.g. A and B are similar items) or versus a complete switch of intent.

## 2 Related Works

Early multi-touch attribution work paved the way for model-based conversion attribution instead of heuristic rules like last click. Shao and Li [1] pioneered this direction with a logistic regression-based model that learns each channel's contribution from data. By employing a bagged logistic regression ensemble, their method achieved stable attribution estimates for each advertising channel while maintaining predictive accuracy. Building on such foundations, more complex machine learning models have been applied. Deep sequence models (RNNs, Transformers) allow attribution to consider the order and context of ad exposures. Several works integrate attention mechanisms to let the model learn which touchpoints are most influential for conversion [2, 3]. For example, a self-attention layer can weight each interaction in a user's journey and produce a probabilistic credit assignment. Kumar et al. showed that attention-based RNN models can indeed distribute credit across multiple touches [3].

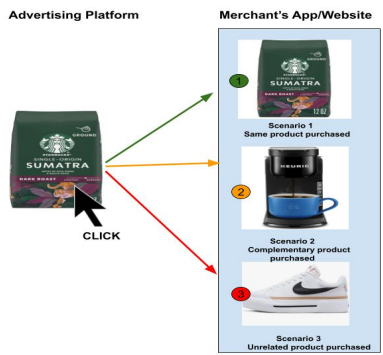

**Figure 1: Showing three different post-click purchase scenarios, Scenario 1 (green) – the same product is purchased (Click A → Buy A, CABA); Scenario 2 (orange) – a complementary product is purchased (meaningful CABB); and Scenario 3 (red) – an unrelated product is purchased (noisy CABB)**

In contrast to deep sequence-based attribution models [2, 3] that leverage RNN's or transformers to learn credit assignments over entire user journeys, we propose a similarity-driven reweighting approach that focuses on product relationships. Our method maps each product to a category (using a taxonomy or category overlap) and employs a collaborative filtering model at the category level to learn pairwise similarity between categories. This yields a standalone attribution weighting mechanism: if a user clicks product A but ultimately purchases product B, the credit assigned to the A click is scaled by the similarity between A's category and B's category. Such a strategy directly addresses the "click A, buy B" problem, ensuring that related product interactions receive due credit even when the exact items differ. Unlike complex sequential neural models which attempt to capture long-range dependencies but often behave as black-boxes, the similarity-based approach is inherently interpretable – the contribution of each click can be explicitly traced to category-level similarity scores. It is also lightweight to deploy, relying on precomputed taxonomy mappings and similarity matrices rather than heavy real-time sequence modeling. Moreover, by abstracting user behavior to the category level, our method is more robust to data sparsity and misalignment in long user buying journeys i.e. even when exact product sequences are infrequent or user paths are lengthy, it can still generalize attribution based on broader item affinities. Multi-Task Learning (MTL) has been widely explored in recommendation and advertising systems to jointly optimize for multiple, often correlated, objectives. General MTL frameworks [4, 5] have laid the groundwork for architectures that can learn shared representations while specializing for distinct tasks. In the context of conversion modeling, a prominent line of work focuses on decomposing post-click outcomes sequentially. For instance, the Entire Space Multi-Task Model (ESMM) [6] effectively estimates post-click conversion rate by modeling both click-through rate and click-through conversion rate, thereby addressing data sparsity and sample selection bias.

To the best of our knowledge, no prior attribution or multi-task conversion model explicitly disentangles Click A → Buy B events or learns to trust them only when the purchased item is semantically related to the clicked one. Consequently, our CABA–CABB decomposition with taxonomy-aware similarity weighting constitutes the first end-to-end framework that directly models and denoises cross-item conversions in large-scale e-commerce recommendations

## 3 Proposed Approach

## 3.1 Problem Formulation as Multitask Learning

We formulate the recommendation conversion prediction as a multitask learning problem with two related objectives:

- **Task 1: CABA prediction (Click A → Buy A)** – predict the probability that a user will purchase the same product they clicked.
- **Task 2: CABB prediction (Click A → Buy B)** – predict the probability that a user's click on product A will result in a purchase of a different product B.

We train a single deep neural network model with shared features and representations i.e. embedding layers but has separate output layers and loss terms for each task. Multitask learning allows shared knowledge transfer between tasks (Figure 2). For example, signals that strongly indicate high user intent could boost both the probability of CABA and CABB, whereas signals of relative dissatisfaction e.g. quick bounce from A's page might raise CABB likelihood and lower CABA likelihood. By training on both outcomes, the model learns a nuanced understanding of post-click behavior that a single-task model would miss.

## 3.2 Training Data and Label Partitioning

The training data for our method is derived from e-commerce session logs, which contain sequences of user interactions (product impressions, clicks, and conversions). We construct labeled examples at the granularity of user, session, product clicked, conversion label. Each such example is associated with two binary labels:

- **$y_1$ (CABA label):** 1 if the user ended up purchasing product A in that session, 0 otherwise.
- **$y_2$ (CABB label):** 1 if the user purchased something in the session other than, 0 otherwise.

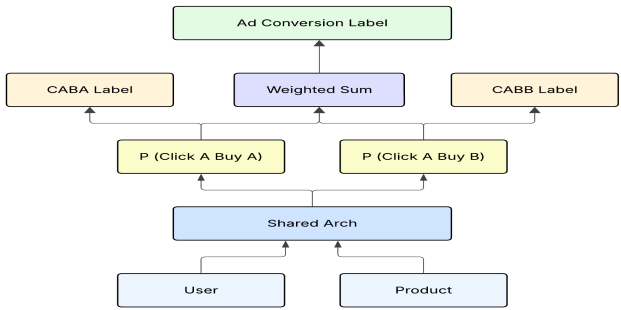

**Figure 2: Two-head multi-task conversion model**

In many cases $y_1 = y_2 = 0$ (no conversion from that click at all). And it's possible (though relatively rare in a short session) that both could be 1 if the user bought multiple items including A and others – in our data preprocessing we treat each purchase event separately. Essentially, we partition conversion events into two buckets: same-item conversions feed the CABA task, and other-item conversions feed the CABB task.

A challenge with this labeling is that the CABB category can be noisy. It treats all instances of click A, buy B ≠ A as equal, even though some B's might be closely related to A while others might be completely unrelated. Training a model to predict a generic "bought something else" probability could be misleading and does not incentivize the model to capture users true shopping intent. To address this, we introduce a similarity measure (next section) between products and use it to weight the CABB training examples.

## 3.3 Category-Level Similarity via Product Taxonomy

We create similarity metric between product ids to weight CABB training examples in the following manner:

- **Taxonomy Mapping**: Each product is mapped to its corresponding leaf node in a static, hierarchical product taxonomy (e.g., Home → Kitchen → Coffee Makers or Grocery → Beverages → Coffee Beans). Anchoring at the leaf level prevents over-generalization, reduces sparsity by pooling engagement across many item ids that share the same fine-grained category, and preserves user intent for close-substitute items within a category.
- **Category Co-Engagement Signals**: Instead of computing similarity directly between individual items, we aggregate user engagement at the leaf category level. Using session logs of user behavior (clicks, add-to-cart actions, purchases, etc.), we measure how often products from category X and category Y are co-engaged within the same sessions. For example, if many users who click on items in Coffee Makers also end up purchasing items in Coffee Beans, this indicates a strong behavioral linkage between these two categories despite their different taxonomy mappings.
- **Category-to-Category Similarity Model**: To learn category-to-category similarity, we adopt a classic item-to-item collaborative-filtering formulation applied at the leaf-category level [7]. First, each category c is embedded in user space as a high-dimensional vector $v_c \in \mathbb{R}^{(|U|)}$ whose $u^{th}$ coordinate accumulates that user's interactions with the category:

$$(v_c)_u = w_{click} * \#clicks + w_{cart} * \#add\_to\_carts + w_{purchase} * \#purchases + \ldots \quad (1)$$

Event-type weights $w$ up-weight higher-intent signals (e.g., purchases) while retaining lower-intent indicators (e.g., clicks) for additional context.

Pairwise cosine similarity between these vectors yields the category-similarity matrix.

$$S(c_i, c_j) = \frac{v_{c_i} \cdot v_{c_j}}{|v_{c_i}|_2 \, |v_{c_j}|_2}, \; S(c_i, c_j) \in [0,1] \quad (2)$$

Because the interaction counts are non-negative, cosine values naturally lie in [0,1]; categories whose items are frequently co-engaged by the same users receive scores approaching 1, whereas unrelated categories trend toward 0. For any product pair $(A, B)$, we then look up their respective leaf categories and assign the example-level weight:

$$\alpha_{(AB)} = S(cat(A), cat(B)) \in [0,1] \quad (3)$$

thereby capturing both strong substitution signals within a category and meaningful cross-category complements exposed through collective user behavior.

### 3.4 Multi-Task Loss with Category-Similarity Weighting

We embed the category-level similarity signal from Section 3.3 directly into the multi-task objective so that the model learns to trust Click A → Buy B (CABB) events only when the clicked and purchased products are weighted significantly by the similarity score. For a batch B of click events, the two task-specific cross-entropy losses are:

$$\mathcal{L}_{\text{CABA}} = -\sum_{(A,B) \in \mathcal{B}} [y_1 \log \widehat{y_1} + (1 - y_1) \log(1 - \widehat{y_1})] \quad (4)$$

$$\mathcal{L}_{\text{CABB}} = -\sum_{(A,B) \in \mathcal{B}} \alpha_{AB} [y_2 \log \widehat{y_2} + (1 - y_2) \log(1 - \widehat{y_2})] \quad (5)$$

The total multi-task objective combines the two with a tunable balance λ:

$$\mathcal{L} = \mathcal{L}_{\text{CABA}} + \lambda \, \mathcal{L}_{\text{CABB}} \quad (6)$$

With this formulation, the model effectively de-biases conversion attribution by grounding cross-item learning in semantic as well as engagement driven category relationship between products.

## 4 Offline Experiments

We conducted extensive experiments on an internal proprietary e-commerce dataset to evaluate the proposed CABB-aware multitask model. The dataset consists of user sessions from our online e-commerce advertising platform, including product impressions, clicks, and purchases for opt-in users only. We withhold specific dataset details for confidentiality but note that it is large-scale and reflective of real-world shopping behavior with abundant CABB instances. Our experimental study is designed to answer four key research questions:

**(RQ-1) Overall Performance**: Does the multitask CABA+CABB model improve overall recommendation performance compared to a baseline that ignores the CABB issue i.e. a standard last-click conversion attribution model?

**(RQ-2) Performance breakdown by CABA vs CABB**: Does our model improve performance of both CABA and CABB tasks?

**(RQ-3) Effect of Embedding Choice**: How does the choice of item embedding influence the model's performance?

**(RQ-4) Feature importance analysis**: How does CABA and CABB tasks differ with respect to important features?

We use Normalized Entropy (NE) to evaluate the ranking quality and calibration of conversion predictions. Normalized entropy is essentially the model's cross-entropy loss normalized by the entropy of the base rateshagunsodhani.com. It is a standard metric in advertising and recommender evaluations for probabilistic predictions [8]; lower NE indicates better performance (with NE = 1 meaning the model is no better than predicting the average conversion rate, and NE = 0 indicating a perfect model).

### 4.1 RQ-1 Overall Performance

To answer RQ-1 we benchmark our final multitask model against two baselines: 1) Baseline 1 – is a classic single task conversion model which attributes all the conversions to the last clicked product 2) Baseline 2 – considers only CABA conversion data to prevent biasing the model towards unrelated conversions.

As shown in Table 1, our final model achieved significantly better overall NE than the baselines. The NE of our final model was 0.495, compared to 0.575 for the single-task Baseline-1(a 13.9% reduction, which is a substantial gain in this metric), and 0.547 for Baseline-2 which only used CABA conversions. These results validate our intuition that decoupling CABA, CABB tasks and using similarity weights to denoise CABB conversions can improve the model. Note that since our model performs better than Baseline-2, it demonstrates that using CABB data with similarity weighting adds incremental value to our final model.

Table 1: Overall NE Comparison

| Test Variant | NE |
|---|---|
| **Baseline 1** | 0.575 |
| **Baseline 2** | 0.547 |
| **Multitask Model** | 0.495 |

### 4.2 RQ-2 Performance Breakdown by CABA vs CABB Tasks

To understand how the multitask formulation balances same-item (CABA) and cross-item (CABB) performance, we report Normalized Entropy for each task while sweeping the weighting coefficient λ in Eq. (6).

As shown in Table 2, the single task last clicks attribution baseline which does not distinguish CABA and CABB conversions performs the worst on both tasks separately.

Removing CABB supervision (λ = 0) improves CABA task but performs badly on CABB task. In our proposed formulation as λ increases, CABB NE improves monotonically from 0.692 → 0.505 while the performance on CABA is relatively the same showing the benefit of weighted cross-item supervision.

These results validate our hypothesis: separating and properly weighting CABB events corrects attribution bias and yields a more balanced predictor than either conventional last-click training or a CABA-only model.

Table 2: Impact of CABB-loss weight λ on task-level NE

| λ (CABB weight) | CABA NE | CABB NE |
|---|---|---|
| Baseline (Single task) | 0.512 | 0.643 |
| 0.0 | 0.398 | 0.692 |
| 0.1 | 0.409 | 0.651 |
| 0.25 | 0.419 | 0.581 |
| 0.50 | 0.432 | 0.550 |
| 0.75 (Selected) | 0.451 | 0.519 |
| 0.9 | 0.479 | 0.505 |

## 4.3 RQ-3 Effect of Embedding Choice

We also compared the effect of the similarity term in Eq. (3) which tells the model which Click A → Buy B events to trust. We evaluate three ways of computing the similarity weight each plugged into the same multitask model (λ = 0.75, identical architecture, data splits, and hyper-parameters) so that any performance difference is attributable solely to the weighting scheme. (1) Static = 1 ("No-Weight") assigns $\alpha_{(AB)}$ = 1 for every click–purchase pair, treating all CABB events as equally informative and thus ignoring product relatedness. (2) similarity from i2i embeddings learnt from co-engagement between products which likely misses the semantic / substitutable relationship (3) Taxonomy + Collaborative Filtering ("Ours") maps each product to a leaf in the merchant's hierarchical taxonomy, then learns a category-to-category similarity matrix from large-scale co-engagement logs, enabling $\alpha_{(AB)}$ to reflect both substitutable (same-category) and complementary (cross-category) relationships grounded in real shopping behavior.

As shown in Table 3, simply adding every CABB example equally ('Static = 1') injects noise: CABA NE improves significantly after de-noising the data via similarity weighting from 0.478 to 0.441. Item level i2i embedding similarity recover some of the gap but likely suffers from sparsity and lack of explicit semantic information. Note that because $\alpha_{(AB)}$ down-weights Click A → Buy B pairs, the model intentionally focuses on a cleaner subset of CABB events. When evaluated uniformly over all CABB instances, this yields a slight increase in CABB NE (Table 3). However, the large CABA gain indicate that discounting noisy cross-item conversions is a worthwhile trade-off.

These results confirm that the way similarity is modelled critically influences how effectively the multitask framework learns from cross-item conversions. A taxonomy-grounded, engagement aware signal captures both substitution and complementarity, unlocking the full benefit of CABB supervision.

Table 3: Effect of Similarity Weighting Schemes on NE

| Test Variant | CABA NE | CABB NE |
|---|---|---|
| Static = 1 (No-Weight) | 0.478 | 0.552 |
| I2I embedding similarity | 0.462 | 0.558 |
| Taxonomy + CF (Ours) | 0.441 | 0.563 |

## 4.4 RQ-4 Feature importance analysis

To understand differences among the top ranked features for CABA and CABB tasks, we ran feature importance analysis for the top ranked features for each of these tasks respectively.

- **CABA Feature Importance** – Most of the top ranked features for CABA tasks were broadly related to personalization e.g. users' past interaction with clicked product or similar products, time gap since the user interacted with clicked product or similar products.
- **CABB Feature Importance** – Most of the top ranked features for CABB tasks were broadly related to semi-personalization (due to similarity weighting) and overall popularity e.g. advertiser historic CVR globally, users' past purchases with same/similar category products. Note that even after similarity weighting, we expect some global popularity features to show up in CABB because similarity metrics is also learnt on co-engagement data which is dominated by popular products.

Qualitative analysis of important features confirmed our intuition that CABA task is influenced much more by features targeting personalization whereas CABB task is influenced by semi personalization-based features due to the similarity weighting scheme used in rebalancing its loss.

## 5  Online Deployment

To validate real-world performance, we deployed our best offline model in a live A/B test on our e-commerce platform's recommendation system. The test ran for two weeks, comparing our new model against the existing production recommender (which was based on single task last-click attribution). The reported improvements in business metrics are statistically significant.

The results in the online setting mirrored the offline gains, demonstrating the practical value of our approach. The treatment delivered a +0.25 % lift in our primary business metric (proprietary metric correlated with the long-term health of our business). Crucially, the live experiment also showed an increase in CABA rate by 1.27% i.e. the proportion of sessions where the user's purchase was from the same initial click rose supporting the claim that our model kept recommendations more personalized and aligned with users shopping intent. From a business perspective, these improvements translate to a better user experience and potentially higher revenue. By mitigating the Click A, Buy B problem, the taxonomy-aware recommender not only avoids the pitfall of misleading recommendations but also capitalizes on genuine cross-sell opportunities (through complementary items). We continue to monitor longer-term impacts such as repeat purchase rate and customer retention, hypothesizing that the more contextually relevant recommendations will foster greater trust in the platform's suggestions over time.

## 6 Limitations

While the proposed taxonomy-aware co-engagement similarity weighting is central to the model's success in distinguishing meaningful CABB events, this approach introduces some limitations. Firstly, the system's efficacy is heavily dependent on the quality and granularity of the underlying product taxonomy; an outdated, poorly structured, or inappropriately granular taxonomy can hinder the accurate identification of substitutable or complementary relationships, thereby degrading the precision of the similarity weights. Secondly, although the co-engagement signals aim to highlight genuine cross-item intent, they can sometimes be noisy. Unrelated products that happen to co-occur frequently in user sessions due to general popularity or some seasonal trends might still acquire non-negligible similarity scores. This can lead to the misattribution of importance to some unrelated CABB events, introducing subtle noise into the learning process. Finally, the multi-task framework involves careful calibration of the loss contributions from the CABA and CABB tasks using the hyperparameter λ. Determining an optimal λ typically requires extensive empirical tuning, and the model's performance can be sensitive to this value, potentially requiring re-tuning as user behavior or product catalogs evolve.

## 7 Conclusion and Future Work

The prevalent 'Click A, Buy B' (CABB) phenomenon in e-commerce poses a significant challenge to traditional conversion attribution models, often leading to biased learning and suboptimal recommendations. This paper introduced a novel approach to address this by reframing conversion prediction as a multi-task learning problem, with distinct heads CABA and CABB conversions. A key contribution is our taxonomy-aware collaborative filtering weighting scheme, which aims to discern meaningful CABB events (substitutions or complements) from coincidental purchases by leveraging category-level co-engagement signals.

Our offline experiments demonstrated a 13% reduction in normalized entropy compared to a last-click baseline, and our model outperformed a CABA-only baseline, underscoring the value of carefully incorporating weighted CABB signals. Crucially, online A/B testing on a large-scale e-commerce platform yielded a statistically significant +0.25% lift in our primary business metric and an increase in the CABA rate, indicating more personalized and aligned recommendations.

We acknowledge that the reliance on taxonomy and co-engagement for similarity scoring can be sensitive to taxonomy quality and may not perfectly filter all unrelated CABB events. To address these limitations and further enhance the model, future work could explore several avenues. Incorporating world knowledge from Large Language Models (LLMs) could provide a more robust method for labeling CABB events as related or unrelated, potentially augmenting or refining the co-engagement-based similarity scores. Furthermore, instead of a fixed hyperparameter λ, an attention mechanism could be developed to dynamically learn the optimal balance between CABA and CABB loss contributions, potentially using the final ad conversion event as a supervisory signal. Exploring such dynamic weighting, extending the attribution window beyond single sessions, and integrating richer semantic product features remain promising directions for advancing CABB-aware recommendation systems.